\documentclass{article}

\usepackage{graphics,amsfonts}
\title{Numerical Simulations of Finite Dimensional Spin Glasses Show a 
Mean Field like Behavior}

\author{E Marinari%
  \footnote{Dipartimento di Scienze Fisiche, Universit\`a di Cagliari 
            (Italy),
            e-mail: \texttt{marinari@ca.infn.it}. Invited talk given 
            at the ICMP 97 Brisbane Mathematical Physics Conference}
}

\date{September 10, 1997}

\begin{document}

\maketitle

\abstract{I discuss results from numerical simulations of finite 
dimensional spin glass models, and show that they show all signatures 
of a mean field like behavior, basically coinciding with the one of 
the Parisi solution. I discuss the Binder cumulant, the probability 
distribution of the order parameter, the non self-averaging behavior. The 
determination of correlation function and of spatially blocked 
observables quantities helps in qualifying the behavior of the system.}

\section{Introduction\protect\label{S-INTROD}}

In the following I will try to introduce the main results that come 
from numerical simulations of finite dimensional spin glass models. 
In this short report I will just try to summarize the main ideas, the 
crucial points and their implications. The reader is referred to the 
original papers for details. The recent review \cite{OUR-REVIEW} is 
a detailed summary of the history of numerical simulations in the field. 
References \cite{MAPARI,MPRR,KAWYOU} establish and qualify the 
existence of a phase transition in the $3$ dimensional model. 
Reference \cite{CAMAPA} discusses about ultrametricity in the finite 
dimensional models, while \cite{PICRIT,MAPAZU} discuss about the 
existence of a transition in field. Reference \cite{IMPR} discusses 
in detail about the probability distribution of the overlap in $3$ 
dimensional models, while \cite{MAPARU} and \cite{MARZUL} are 
respectively accurate simulations of the $3$ and of the $4$ 
dimensional model. Improved Monte Carlo methods like {\em tempering}, 
that turn out to be crucial in allowing interesting numerical 
simulations of the disordered phase of spin glasses, are discussed in 
\cite{METHODS}. These numerical results can be a good support in the 
rigorous approach to models with quenched disorder \cite{RIGOR} that 
has been recently moving the first steps.

We are interested in models with quenched disorder.  The Hamiltonian 
$H=-\sum'\sigma_{i}J_{i,j}\sigma_{j}$, where $\sigma_{i}=\pm 1$, the 
indices $i$ and $j$ take values on a $D$ dimensional lattice of linear 
size $L$ (typically with periodic boundary conditions), and the primed 
sum is over first neighboring site couples on a simple cubic lattice.  
The {\em realistic} Edwards-Anderson model is defined in $3D$ (even 
if, as we will see in the following, we are mainly interested in the 
generic behavior of finite dimensional models, as opposed to the 
infinite dimensional mean field approximation, and sometimes we will 
prefer to study the $4D$ model, not to be mislead by the additional 
difficulties that the vicinity of $D=3$ to the lower critical 
dimension of the model can introduce).  The couplings $J$ are quenched 
random variables: they can, for example, take the value $\pm 1$ with 
probability one half, or be Gaussian, with $P(J) \approx 
\exp{(-J^{2})}$.  The crucial fact is that they are fixed and that 
they create a random frustration \cite{THE-BOOK}.

Since this model cannot be solved (obviously) one studies its mean 
field version, the Sherrington Kirkpatrick model (SK).  The sum 
defining the Hamiltonian runs now over all site couples of the 
lattice, i.e.  $D=\infty$: to make the infinite volume limit well 
defined the $J$ variables have to scale now like $N^{-\frac12}$.  Also 
the mean field theory is highly non-trivial.  What is currently 
believed to be the correct solution of the model has been given by 
Parisi \cite{PARISI} and relies on the mechanism of the so called {\em 
Replica Symmetry Breaking} \cite{THE-BOOK}.

The order parameter is the {\em overlap} $q\equiv 1/V 
\sum_{i}\sigma_{i}\tau_{i}$ of two configurations of the system in the 
same realization of the quenched disorder, $J$. For a given $J$ we 
call $P_{J}(q)$ the probability distribution of $q$, and by defining 
with a bar the average over the quenched disorder we call 
$P(q)\equiv\overline{P_{J}(q)}$. All the observables we will discuss 
in the following (like for example the susceptibility) will be similar 
to what we would study in the case of a normal spin model, 
substituting to the magnetization $m$ the two replica overlap $q$.

Parisi solution of the mean field theory of spin glasses has many new 
and remarkable features \cite{THE-BOOK}, that make it a new paradigm.  
Let us just quote in short some of the most relevant ones.  A 
non-trivial equilibrium distribution of the order parameter denotes 
the existence of many states non related by a simple explicit 
symmetry.  There is a complex free energy landscape (and all 
temperatures $T$ smaller than the critical value $T_{c}$ are 
critical).  There exist observable quantities (based on $2$ or more 
replicas) that are {\em non-self-averaging}: macroscopic fluctuations 
survive in the infinite volume limit.  The structure of equilibrium 
states is embedded with an ultrametric structure.  The phase 
transition survives the presence of a finite magnetic field.
It is also important to note that this picture, and the structure of 
Replica Symmetry Breaking, could be relevant to the description of 
glasses \cite{VETRI}: at least at the mean field level
a complex pattern of frustration, even without 
quenched disorder, can be enough to put the system in a Parisi-like 
phase \cite{VETRI}.

Here we will discuss results from numerical simulations, done by using 
Monte Carlo and improved Monte Carlo methods \cite{METHODS}.  Field 
theory is applied to the problem, and is starting to give interesting 
hints, but like for numerical simulations it is difficult to get 
reliable results too.  We will quote here our main evidences, that we 
will discuss in some more detail later on, and we refer the reader 
to the original papers for more details.  The $4D$ model (for example) 
turns out to behave very similarly to the mean field model (and, given 
some differences due to the vicinity of the lower critical dimension, 
also the $3D$ model does)\footnote{The upper critical dimension of the 
problem is $6$.}.  This statement is quantitative: one finds for example 
very similar exponents for finite size corrections.  I will recall 
here, in all generality, that since we are talking about numerical 
results, albeit well controlled, one is always dealing with finite 
volume, finite precision, results (with periodic boundary conditions 
in all the following).

There is a clear phase transition with as an order parameter the 
overlap $q$.  One can determine critical exponents with good 
precision.  The $P_{J}(q)$ and $P(q)$ that we have defined before turn 
out to be non-trivial, and we are able to control their infinite 
volume limit.

The analysis of spatial correlation functions shows that the phase 
transition is mean field like: signatures of ultrametricity are 
clearly detected. Again, the behavior of finite $D$ corrections in 
$4D$, for example, is very similar to the one of the mean field 
solution. 

\section{Existence of a Phase Transition\protect\label{S-PHASET}}

\subsection{The Binder Cumulant\protect\label{SS-BINDER}}

The simplest way to qualify the distribution probability of the order 
parameter is based on the use of the Binder cumulant

\begin{equation}
	B_{L}(T)\equiv \frac12
	\left( 
	  3-
	  \frac
	  {\overline{\langle q^{4}\rangle}_{L}(T)}
	  {\left(\overline{\langle q^{2}\rangle}_{L}\right)^{2}(T)}
	\right)\ .
	\label{E-BINDER}
\end{equation}
The shape of $P(q)$ at $T_{c}$ is universal, and 
$B_{c}\equiv\lim_{L\to\infty}B_{L}(T_{c})$ is universal. In the warm 
phase the distribution probability of the order 
parameter is Gaussian and $\lim_{L\to\infty}B_{L}(T>T_{c})=0$, while for 
ferromagnetic spin models in the cold phase 
$\lim_{L\to\infty}B_{L}(T>T_{c})=1$ (here $g$ is computed from $m$ not 
from $q$). Curves of $B(T)$ for different $L$ values cross at $T_{c}$, 
giving a signature of the finite volume pseudo critical point.

In the broken phase of the mean field SK spin glass $g$ is not one for 
$T<T_{c}$.  Here one can compute analytically that the non-trivial 
structure of $P(q)$ implies that $B(T<T_{c})=\frac32-\frac12 \int 
dP(q)q^{4} / (\int dP(q)q^{2})^{2}$.  $g$ is here in the broken phase 
a non-trivial function that is $1$ at $T=0$, has a minimum and tends 
to $1$ when $T\to T_{c}^{-}$.

\begin{figure}
  \begin{center}
    \scalebox{0.75}{\includegraphics{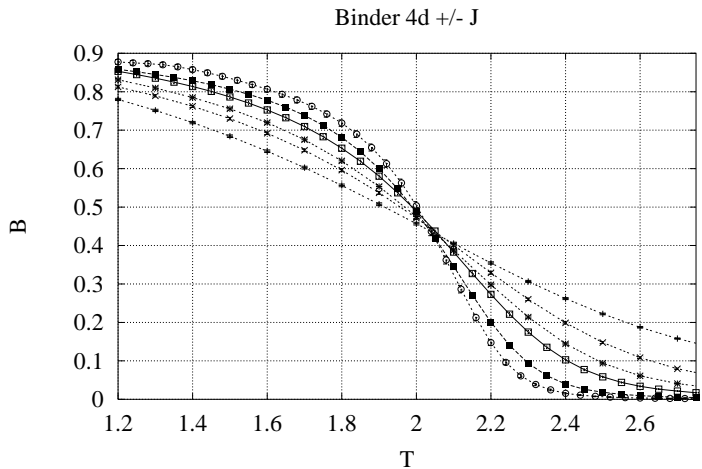}}
    \caption{The Binder cumulant versus $T$.}
    \protect\label{F-BINDER}
  \end{center}
\end{figure}

In figure (\ref{F-BINDER}) (from ref.  \cite{MARZUL}) the Binder 
cumulant for the $4D$ model with binary couplings.  $L$ goes from $3$ 
(lower curve for low $T$) to $10$ (upper curve for low $T$) (we have 
$L=3,4,5,6,8,10$).  The crossing is very clear, and the critical point 
can be found with good precision: the quality of this evidence is 
comparable to the numerical evidence one has in the case of the usual 
ferromagnetic transition for the Ising model.  A scaling plot of 
$B((T-T_{c})L^{1/\nu})$ shows a very good scaling with $T_{c}=2.04$ 
and $\nu=0.91$ (see \cite{MARZUL} for a detailed analysis including 
precise error bars on these numbers). One finds the same kind of good 
scaling for the overlap susceptibility, and determines \cite{MARZUL} 
$\eta=-0.45$ and $\gamma =0.23$ (always in $D=4$).

\subsection{The $P(q)$\protect\label{SS-PQ}}

It is interesting to discuss directly the shape of the probability 
distribution of the order parameter, $P(q)$.

\begin{figure}
  \begin{center}
    \scalebox{0.5}{\rotatebox{270}{\includegraphics{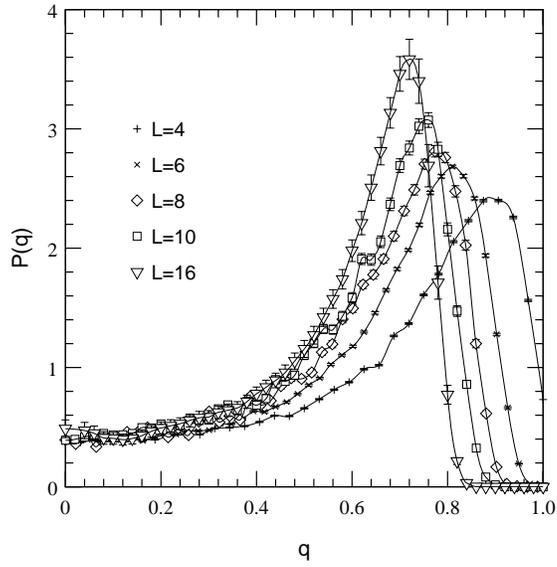}}}
    \caption{Probability distribution of the overlap $P(q)$ for the $3D$
    Ising spin glass with Gaussian couplings. 
    Right to left $L=4,6,8,10$ and $16$ and $T=0.7$.}
    \protect\label{F-PQ}
  \end{center}
\end{figure}

In figure (\ref{F-PQ}) (from ref.  \cite{IMPR}) I show $P(q)$ for the 
$3D$ Ising spin glass with Gaussian couplings for different lattice 
sizes.  Here $T=0.7$, in the cold phase.  Since we are interested in 
the infinite volume limit we have to look at what happens for 
increasing lattice size.  There are two issues that need to be 
discussed.  The first point is about the region of $P(q)$ for small 
$q$: a non-zero probability means here that there are stable states 
with zero overlap.  We notice that when increasing the lattice size 
the non-zero value of the plateau does not change (it would decrease 
if the transition would become of the ferromagnetic type, and it is 
shown in \cite{IMPR} that it does increase in a Kosterlitz-Thouless 
like transition). So, the constant value of the low $q$ plateau is an 
indication of a non-trivial phase in the infinite volume limit.

The second point is about the location of the peak, that should 
converge, in the infinite volume limit, to the value Edwards-Anderson 
value $q_{EA}$ \cite{THE-BOOK}.  The location of the peak shifts 
strongly when increasing the lattice size (we know from a number of 
numerical experiments that finite size effects are very strong in spin 
glasses). It is essential to show that the location of the peak does 
not shift down to $q=0$ when $L\to\infty$. 

\begin{figure}
  \begin{center}
    \scalebox{0.75}{\includegraphics{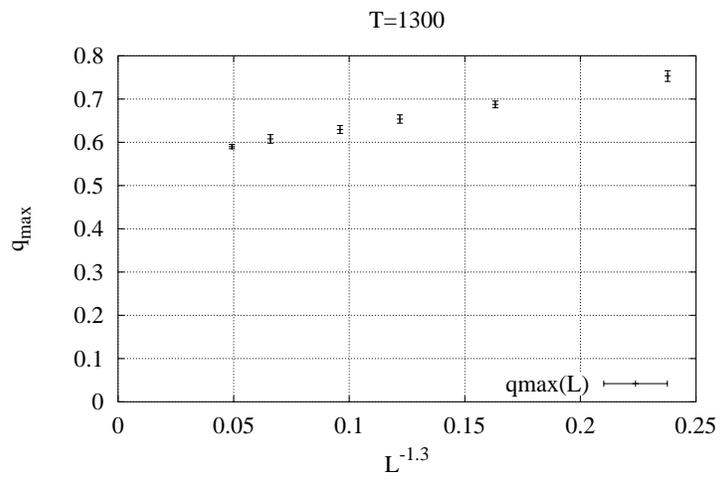}}
    \caption{$q_{max}$ versus $L^{-1.3}$ for the $4D$ Ising spin 
    glass with binary couplings.}
    \protect\label{F-QMAX}
  \end{center}
\end{figure}

This can be shown with high confidence level in $4D$ \cite{MARZUL} 
(and again quite clearly but with a larger uncertainty also in $3D$ 
\cite{IMPR,MAPARU}). In figure (\ref{F-QMAX}) we show the location of 
the peak (for the $4D$ spin glass with binary couplings) at low $T$ 
(in the broken phase) as a function of $L^{-1.3}$ (the power is hinted 
by our best fit, see later). 

Already from figure (\ref{F-QMAX}) one would consider not plausible a 
null limiting value, but one has to be careful, and trying to keep the 
best fit under strict quantitative control.  To be more clear one has 
to exclude, for example, that a fit of the form $A/L^{\alpha}$ could 
work.  This is usually quite tricky, since in a finite $x$ range a 
small negative power mimics quite well a constant value with 
corrections given by a higher negative power.  In $4D$ even this 
evidence starts to be quite good \cite{MARZUL}: one finds that 
$q_{max}=0.55+0.86L^{-1.31}$.  This is a remarkable piece of 
quantitative agreement with the mean field theory: the analytic 
computation of ref.  \cite{FRPAVI} gives for the mean field an 
exponent of $1.25$ (that is very similar to the $1.31$ we find in 
$4D$).

To end this section it is also interesting to note that more and more, 
when increasing $L$, the individual $P_{J}(q)$ (for a given 
realization of the couplings) show a complex structure and many 
minima.  For example \cite{MARZUL} the density of $J$ configurations 
whose  $P_{J}(q)$ has $4$ maxima goes from $0.1$ at $L=3$ up to $.35$ 
for $L=8$.

\section{The Mean Field Picture\protect\label{S-MEANFI}}

\subsection{Non Self-Averageness\protect\label{SS-NONSEL}}

In the mean field picture one expects that sample to sample 
fluctuations of some observable quantities do not go to zero in the 
thermodynamic limit. We have shown numerically in \cite{MPRR} that for 
example sample to sample fluctuations of $\langle q^{2} \rangle_{J}$ 
behave in the $3D$ spin glass as one would expect for the mean field. 
In this same paper we have also noticed (and the same holds for the 
$4D$ model) that the relation that had been obtained in Mean Field 
\cite{THE-BOOK}

 \begin{equation}
 	\overline{\langle q^{2} \rangle_{J} \langle q^{2} \rangle_{J}}
 	= \frac23 
 	\overline{\langle q^{2} \rangle_{J}} \,\,\,\,
 	\overline{\langle q^{2} \rangle_{J}}
 	+ \frac13 \overline{\langle q^{4} \rangle_{J}}\ ,
 	\label{E-GUERRA}
 \end{equation}
holds with very good precision for the finite dimensional model.  
Guerra in ref.  \cite{RIGOR} has shown that indeed this class of 
relations has to hold, also for finite dimensional models, under very 
general hypothesis.

\subsection{Correlation Functions and Block Overlap\protect\label{SS-CORREL}}

To establish the pattern of the symmetry breaking one wants to learn 
about the spatial structure of typical equilibrium configurations. We 
want to learn about the structure of typical spin domains.

One way to study this problem is to look at spatial correlation 
functions.  Mean field theory tells us that, as we have seen (and 
found also in the realistic models), the theory has an equilibrium $q=0$ 
sector.  Here one expects a power behavior, i.e.  the restricted 
$q$-$q$ correlation functions behave as

\begin{equation}
	\frac{1}{V} \overline{\sum_{i}\langle q_{i}q_{i+x}\rangle}|_{q=0}
	\approx |x|^{-\alpha}\ .
	\label{E-CORFUN}
\end{equation}
We have used a dynamic approach to compute these equilibrium 
correlation functions \cite{MPRR}. 
We reach a good control, and we observe a clear 
power behavior. Recently we have confirmed this result with equilibrium 
simulations \cite{MAPARU}.

The second check is based on defining the overlap of two 
configurations in boxes of linear size $R$, $q_{R}(x)\equiv 
R^{-D}\sum_{y}\sigma_{x+y}\tau_{x+y}$, where $y$ is a vector in a 
$R^{D}$ box, and checking how does this observable behave when $R$ 
increases.  In an ordinary scaling picture after reaching the typical 
size of a cluster (for $R\to\infty$, $R\ll L$) the distribution 
probability of $q_{R}$, $P_{R}(q_{R})$, should peak in two 
$\delta$-functions at $\pm q_{EA}$.  On the contrary in the mean field 
approach one expects the $P_{R}$ to stay Gaussian on all scales.  In 
our numerical simulations we always observe a clear mean field like 
behavior.

\subsection{Ultrametricity\protect\label{SS-ULTRAM}}

Ultrametricity is a crucial feature of the Parisi solution (see for 
example \cite{RATOVI}).  States appearing in the Parisi solution of 
the Mean Field SK theory are organized according to an ultrametric 
structure: if we define a distance among two configurations 
$d(\alpha,\beta)$ the usual triangular inequality $d(\alpha,\gamma) 
\le d(\alpha,\beta)+d(\beta,\gamma)$ is substituted by the stronger 
inequality $d(\alpha,\gamma) \le \max \{d (\alpha,\beta) , 
d(\beta,\gamma)\}$, i.e.  all triangles have two equal sides longer or 
equal to the third one. This is the structure one finds when 
considering, for example a hierarchical tree. One possible definition 
for the distance of two spin configurations is

\begin{equation}
   d(\alpha,\beta) \equiv \frac{q_{EA}-q_{\alpha\beta}}{2q_{EA}}\ .
	\label{E-DISTAN}
\end{equation}
To study this issue (in $4D$ with binary coupling) \cite{CAMAPA} we 
have used a constrained Monte Carlo procedure. We simulate $3$ copies 
of the system, with the same couplings, and we constrain $2$ of the 
$3$ overlap to be fix 

\begin{equation}
	q(\alpha,\beta)=q(\beta,\gamma)=\overline{q}
	\label{E-FIXED}
\end{equation}
(in the actual numerical simulation we let $\overline{q}$ vary over a 
small range). We measure then $q(\alpha,\gamma)$.

Again, the results of the simulation are not distinguishable from the 
ones one gets for the SK model, and contains strong hints towards the 
presence of an ultrametric structure. For example we can give a 
measure of the amount of equilibrium configurations that are not 
ultrametric by defining

\begin{equation}
	I_{L}\equiv 
	\int_{-1}^{\overline{q}}(q_{L}-\overline{q})^{2}P(q_{L})dq_{L}
	+
	\int_{q_{EA}}^{1}(q_{L}-q_{EA})^{2}P(q_{L})dq_{L}\ ,
	\label{E-INTEGR}
\end{equation}
that we expect to go to zero, for $L\to\infty$, if the states have an 
ultrametric structure. We find that $I_{L}$ has a very clear power 
behavior, i.e. $I_{L}\approx (-1\pm 5)10^{-4}+(.76\pm 
.03)/L^{2.21\pm 0.04}$ (where the errors are only statistical). 
This is again a remarkable agreement with mean field predictions, 
since from ref. \cite{FRPAVI} one would expect a value of 8/3 for the 
mean field theory.

\section{Conclusions\protect\label{S-CONCLU}}

Numerical simulations show that the behavior of finite dimensional 
spin glasses is very similar to the one of the Parisi solution of mean 
field theory.  Together with a qualitative evidence (including the 
behavior of $P(q)$ and the correlation function) there is some 
striking quantitative evidence, involving the value of exponents 
determining finite size corrections.  It is maybe worth noticing that 
where the numerical evidence is not very strong (because of finite 
size effects or because of the difficulty of thermalizing the system 
deep in the cold phase) this is true also for actual simulations of 
the mean field theory.  There is, as usual, space for improvements.

\section*{Acknowledgments}
Discussions with G. Parisi, J. Ruiz-Lorenzo and F. Zuliani have been 
crucial in drawing the picture I have described here.  I warmly 
acknowledge them.


\begin{thebibliography}{99}

\bibitem{OUR-REVIEW}
E. Marinari, G. Parisi and J. Ruiz-Lorenzo,
{\em Numerical Simulations of Spin Glass Systems,}
in {\em Spin Glasses and Random Fields,}
edited by P. Young, to be published,
cond-mat/9701016.

\bibitem{MAPARI}
E. Marinari, G. Parisi and F. Ritort, 
J. Phys. A: Math. Gen. {\bf 27} (1994) 2687.

\bibitem{MPRR}
E. Marinari, G. Parisi, F. Ritort and J. Ruiz-Lorenzo,
Phys. Rev. Lett. {\bf 76} (1996) 843.

\bibitem{KAWYOU}
N. Kawashima and A. P. Young,
Phys Rev. B {\bf 53} (1996) 484.

\bibitem{CAMAPA}
A. Cacciuto, E. Marinari and G. Parisi,
cond-mat/9608161,
to be published in J. Phys. A: Math. Gen..

\bibitem{PICRIT}
M. Picco and F. Ritort,
cond-mat/9702041.

\bibitem{MAPAZU}
E. Marinari, G. Parisi and F. Zuliani, 
cond-mat/9703253.

\bibitem{IMPR}
D. Iniguez, E. Marinari, G. Parisi and J. Ruiz-Lorenzo,
cond-mat/9707050,
to be published in J. Phys. A: Math. Gen..

\bibitem{MAPARU}
E. Marinari, G. Parisi and J. Ruiz-Lorenzo,
{\em The $3D$ Ising Spin Glass On and Off-Equilibrium,}
to be published.

\bibitem{MARZUL}
E. Marinari and F. Zuliani,
{\em $4D$ Ising Spin Glass,}
to be published.

\bibitem{METHODS}
E. Marinari,
{\em Optimized Monte Carlo Methods,}
lectures given at the 1996 Budapest Summer School on 
{\em Monte Carlo Methods}, to be published by Springer-Verlag, J. Kertesz
and I. Kondor editors,
cond-mat/9612010.

\bibitem{RIGOR}
C. M. Newman and D. L. Stein, Phys. Rev. Lett. {\bf 76} (1996) 515; 
G. Parisi, cond-mat/9603001; C. M. Newman and D. L. Stein, 
adap-org/9603001; cond-mat/9612097; F. Guerra, Int. J. 
Mod. Phys. B {\bf 10} (1996) 1675; C. M. Newman in these proceedings.

\bibitem{THE-BOOK}
M. M\'ezard, G. Parisi and M. A. Virasoro,
{\em Spin Glass Theory and Beyond,}
(World Scientific, Singapore 1987).

\bibitem{PARISI}
G. Parisi, Phys. Rev. Lett. {\bf 43} (1979) 1754; J. Phys. A: Math. 
Gen. {\bf 13} (1980) 1101; 1887; L115;  Phys. Rev. Lett. 
{\bf 50} (1983) 1946.

\bibitem{VETRI}
E. Marinari, G. Parisi and F. Ritort, 
J. Phys. A: Math. Gen. {\bf 27} (1994) 7631; {\bf 27} (1994) 7647; 
{\bf 28} (1995) 327; {\bf 28} (1995) 4481.

\bibitem{FRPAVI}
S. Franz, G. Parisi and M. Virasoro,
J. Phys. I (France) {\bf 2} (1992) 1869.

\bibitem{RATOVI}
R. Rammal, G. Toulouse and M. Virasoro, 
Rev. Mod. Phys. {\bf 58} (1986) 765.

\end{thebibliography}
\end{document}